\documentclass[5p]{elsarticle}

\usepackage{graphicx}
\usepackage{amssymb}
\usepackage{amsmath}
\usepackage[section]{placeins}
\usepackage{etoolbox,lineno}
\usepackage{textcomp}
\usepackage{hyperref}
\usepackage{booktabs}
\usepackage{setspace}

\hypersetup{colorlinks=false,linkcolor=false, linktocpage}

\setcitestyle{authoryear}
\setcitestyle{round}
\setcitestyle{semicolon}

\makeatletter
\def\ps@pprintTitle{%
 \let\@oddhead\@empty
 \let\@evenhead\@empty
 \def\@oddfoot{}%
 \let\@evenfoot\@oddfoot}
\makeatother

\begin{document}

\let\today\relax

\begin{frontmatter}

\title{On random number generators and practical market efficiency}

\author[first,second,third,fourth]{Ben Moews\corref{corresponding}}
\ead{ben.moews@ed.ac.uk}
\cortext[corresponding]{Corresponding author}

\address[first]{Business School, University of Edinburgh, 29 Buccleugh Pl, Edinburgh, EH8 9JS, UK}
\address[second]{Centre for Statistics, University of Edinburgh, Peter Guthrie Tait Rd, Edinburgh, EH9 3FD, UK}
\address[third]{McWilliams Center, Carnegie Mellon University, Hamerschlag Dr, Pittsburgh, PA 15213, USA}
\address[fourth]{Pittsburgh Supercomputing Center, 300 S Craig St, Pittsburgh, PA 15213, USA}

\begin{abstract}
Modern mainstream financial theory is underpinned by the efficient market hypothesis, which posits the rapid incorporation of relevant information into asset pricing. Limited prior studies in the operational research literature have investigated tests designed for random number generators to check for these informational efficiencies. Treating binary daily returns as a hardware random number generator analogue, tests of overlapping permutations have indicated that these time series feature idiosyncratic recurrent patterns. Contrary to prior studies, we split our analysis into two streams at the annual and company level, and investigate longer-term efficiency over a larger time frame for Nasdaq-listed public companies to diminish the effects of trading noise and allow the market to realistically digest new information. Our results demonstrate that information efficiency varies across years and reflects large-scale market impacts such as financial crises. We also show the proximity to results of a well-tested pseudo-random number generator, discuss the distinction between theoretical and practical market efficiency, and find that the statistical qualification of stock-separated returns in support of the efficient market hypothesis is dependent on the driving factor of small inefficient subsets that skew market assessments.
\end{abstract}


\end{frontmatter}

\nolinenumbers

\section{Introduction}
\label{sec:introduction}

\let\thefootnote\relax\footnotetext{The Version of Record of this manuscript has been published and is freely available in Journal of the Operational Research Society, 2023-07-20, \href{https://doi.org/10.1080/01605682.2023.2219292}{https://doi.org/10.1080/01605682.2023.2219292}.}One of the primary constituents of financial research is the efficient market hypothesis, which, depending on its variation, prohibits the possibility of significant forecasting based on different kinds of data due to the sufficiently fast incorporation of available information into asset prices \citep{Fama1965, Fama1970}. It is, in some settings, linked to the hypothesis that markets are inherently unpredictable due to following random walks to varying degrees, effectively viewing financial time series as martingales or submartingales \citep{Kendall1953, Cootner1964, Malkiel1973}.

Randomised algorithms employing a random number generator (RNG) are ubiquitous in research applications, including fields as diverse as politics, biology, and cosmology \citep[see, for example,][]{Carson2009, Chaudhary2014, Moews2019a}. The most common application area is, of course, cryptography, as encryptions that underlie electronic communication protocols and, by extension, the Internet, rely on hard-to-predict pseudo-RNGs \citep{Cavusoglu2016}. Given these security challenges, there was an early desire to develop statistical tests for randomness, most famously the Diehard Battery of Tests, which contains the overlapping permutations test applicable to binary sequences \citep{Marsaglia2002}.

In the literature on financial machine learning, a common way to approach challenges into a two-class forecasting problem is to transform datasets to a binary representation \citep{Fischer2018, Lee2019, Moews2019b}. When doing so in a way that removes known market features such as heteroskedasticity, meaning a lack of variance homogeneity along the evolution of a given time series, we can pose the question whether tests assessing the quality of RNGs can then be applied to investigations of market efficiency.

The collection of prior research features two works covering this point of view, both in the operational research literature. First, \citet{Doyle2013} introduce the application of the overlapping permutations test to the efficient market hypothesis in an exploratory study, analysing daily closing prices for 76 broad exchange indices and finding non-uniformity of changes in returns for a subset of them.

Explicitly building on the latter study, \citet{Noakes2016} then focus on the Johannesburg Stock Exchange to investigate the efficiency of small, mid, and large market capitalisation indices over the 2005--2019 period. They extend the mentioned prior research by including adjustments for thin trading, meaning periods of no or low trading volumes, due to the same use of daily price series, and find more evidence for inefficiency among indices for companies with small market capitalisations.

In this paper, we confirm the viability of cross-disciplinary methodology transfers, from the field of random number generation to econometrics, bridging the gap through the application of operational research to the study of financial markets. We combine the strengths of the two existing studies in the literature by focussing on a single exchange generally considered to be efficient in the financial literature, and spanning a both larger and more recent time frame. We also make use of both monthly and daily returns, deviating from previous works by studying market efficiency over longer time horizons made available for information incorporation.

The further contributions of this paper are fourfold and go beyond the scope of the above-mentioned prior research. First, we investigate two sets of experiments, one separated by years and one by companies, to quantify variations in efficiency for both variables, and verify considerable annual variations. We challenge the latter finding with an analysis of cross-correlation systematics through monthly distributional sums, in which the impact of the recent global financial crisis can be observed.

Next, we compare both types of experiments to a state-of-the-art pseudo-RNG as a baseline for the overlapping permutations test, and find that company-separated tests show statistically significant inefficiencies by a slim margin, while year-separated tests paint a clearer picture of a lack of market efficiency. We then consider the role of a small subset of inefficiently traded outliers, demonstrating that company-separated return series fully qualify for randomness under the given test with only a small percentage of companies omitted, and put this finding in the context of prior results.

Lastly, we discuss the notions of theoretical and practical market efficiency as well as consequences of the former, and describe the sufficiency of our results for the latter. Our results have implications for the application of cryptographic tests in financial research, the evolution of weak-form inefficiency as an anomaly on volatile exchanges in developed markets, and the study of exchange inefficiency on the firm level.

\section{Theory, data, and methodology}
\label{sec:theory_data_and_methodology}

\subsection{Information efficiency in financial markets}
\label{sec:information_efficiency_in_financial_markets}

As one of the cornerstones of modern financial theory, the efficient market hypothesis (EMH) makes statements about the incorporation of relevant information into stock prices. Initially proposed by \citet{Fama1965}, it branches into three major variations:
\begin{itemize}
\item The strong form states that asset prices reflect all information, both public and private, due to a timely incorporation by market participants.
\item The semi-strong form relaxes this position and states the above only for publicly available information, allowing for profitable insider trading.
\item The weak form, in a further constriction, posits that asset prices reflect past stock market information such as prices and trading volumes.
\end{itemize}
The weak-form EMH is of special interest for us, as it concerns the incorporation of past information regarding stock behaviour into the market, as opposed to newly emerging information such as earnings announcements. The latter can, due to the randomness of unpredictable new information, be viewed as noise injections into the market in the context of time series of returns, whereas past stock information should not have a significant impact on future market performance under the umbrella of all forms. While the prior literature on the topic of this paper does not cover market efficiency beyond the above, it is useful to provide a short overview. \citet{Fama1970} frames the hypothesis in terms of expected returns,
\begin{equation}
\mathbb{E} ( \tilde{p}_{i, t + 1} | \Phi_t) = [1 + \mathbb{E} (\tilde{r}_{i, t + 1} | \Phi_t)] p_{i, t},
\label{eq:emh_basis}
\end{equation}
with $p_{i, t}$ as the price of a given security $i$ at time $t$, and accordingly for $p_{i, t + 1}$, whereas $r_{i, t}$ denotes the return percentage, meaning $r_{i, t} = (p_{i, t + 1} - p_{i, t}) / p_{i, t}$. $\Phi_t$ represents information assumed to be incorporated into $p_{i, t}$, and the tilde operator signifies the role as a random variable.

This formulation, despite its widespread adoption in financial economics, has not met universal approval. An early criticism is made shortly after by \citet{LeRoy1976}, who describes the definitions used in \citet{Fama1970} as tautologies, an assessment repeated later as ``[...] applying a conditional expectations operator to the identity defining the rate of return as equal to the price relative $p_{t + 1} / p_t$ (less one).'' \citep{LeRoy1989}. Following \citet{Fama1970}, the position that $p_{i, t}$ fully reflects $\Phi_t$ then implies that
\begin{equation}
\begin{split}
\mathbb{E} (\tilde{\alpha}_{i, t + 1} | \Phi_t) &= 0 \mathrm{, with \ } \\
\alpha_{i, t + 1} &= p_{i, t + 1} - \mathbb{E} (p_{i, t + 1} | \Phi_t).
\end{split}
\label{eq:emh_alpha}
\end{equation}
The same holds for returns, meaning
\begin{equation}
\begin{split}
\mathbb{E} (\tilde{\beta}_{i, t + 1} | \Phi_t) &= 0 \mathrm{, with \ } \\
\beta_{i, t + 1} &= r_{i, t + 1} - \mathbb{E} (r_{i, t + 1} | \Phi_t).
\end{split}
\end{equation}
This is generally referred to as a ``fair game'' with respect to the available information by \citet{Fama1970}. As for Equation~\ref{eq:emh_basis}, \citet{LeRoy1989} criticises that these equations follow from the definitions of $\alpha_{i, t + 1}$ and $\beta_{i, t + 1}$ with expectations conditional on $\Phi_t$ on both sides, and argues that the former two definitions as fair game variables do not restrict the stochastic process of the price. The implication is that any capital market would be efficient, while no empirical data could decide on market efficiency.

 Later alternatives to these definitions include the reference to a true price model for assessing equilibrium values available to market agents, although this is acknowledged to introduce a joint hypothesis problem by \citet{Fama1991}, and these definitions continue to face criticisms as tautological \citep{Pilkington2016}. While the purpose of this section is a short-form overview of the background and equations commonly encountered, these objections should be kept in mind when assessing the literature, and reviews from different perspectives are available to the interested reader \citep{Malkiel2005, LeRoy2010, Ausloos2016}. Under the assumption that
\begin{equation}
\begin{split}
\forall t \forall \Phi_t : \mathbb{E} (\tilde{p}_{i, t + 1} | \Phi_t) &\geq p_{i, t} \\
\Rightarrow \forall t \forall \Phi_t : \mathbb{E} (\tilde{r}_{i, t + 1} | \Phi_t) &\geq 0,
\end{split}
\end{equation}
the time series of prices $\{p_{i, t} \}$ follows a submartingale. Interpreting market efficiency as the independence of successive returns, an additional assumption can be made, which is their identical distribution. This leads, as conditional and marginal probability distributions of independent random variables are identical, to
\begin{equation}
f(r_{i, t + 1} | \Phi_t) = f(r_{i, t + 1}),
\label{eq:random_walk}
\end{equation}
for a density function $f$ that is invariant to $t$. While widely accepted in mainstream financial theory, the EMH has attracted criticisms from the field of behavioural economics early on, for example by \citet{Nicholson1968}, and the general counterargument can be summarised as the doubtful statement that, maybe, people are not quite as rational as the mathematical maximisation of utility functions seems to imply \citep{DellaVigna2009}. These criticisms from a behavioural perspective persist until today, with a recent review available in \citet{Kapoor2017}.

In more recent times, the field has further expanded into findings from neuroscience, with corresponding attacks on orthodox market efficiency \citep{Ardalan2018}. Despite this, the hypothesis has proven to possess explanatory power, thus cementing its place in the literature, and the results in this paper paint a picture of explicable variations rather than its rejection from a practical perspective.

While this section targets a limited overview, one alternative is of special interest in the discussion of Section~\ref{sec:discussion} and warrants a short introduction due to its place between the EMH and behavioural criticisms mentioned above. Introduced by \citet{Lo2004} the adaptive market hypothesis aims to reconcile the dominant notion of market efficiency with the findings of behavioural finance from an evolutionary perspective.

Contrary to the assumption that market forces are strong enough to overcome behavioural biases in aggregate, this alternative argues based on bounded rationality as pioneered by \citet{Simon1955}, as opposed to the axiom of rational expectations. Using this framework's assumption of ``satisficing'', the adoption of satisfactory choices due to the costs and limitations of human decision-making, the latter is explained through heuristics that are developed and adapted in an evolutionary learning process.

Should market circumstances change, maladaptive heuristics grow to be unfit, and market actors' behaviour needs to change to remain competitive. Changes in market efficiency in this context can be described, in simple terms, as markets being more efficient if many market agent ``species'' compete for limited financial opportunity resources, as opposed to few species competing for abundant resources.

The adaptive market hypothesis has found empirical success in the analysis of United States stock markets \citep{Urqhart2014}. Other studies from the last few years cover European and Asiant markets, as well as cryptocurrency exchanges. \citep{Urqhart2016, Chu2019, Xiong2019}. A recent overview for the interested reader can be found in \citet{Lo2017}. With this short primer covered, we can now think about the implications for binarised series of stock market returns and their relationship to random number generation.

\subsection{Exchange and empirical data description}
\label{sec:exchange_and_empirical_data_description}

We retrieve monthly close prices of Nasdaq-listed stocks, spanning the years 2001--2019, from the Wharton Research Data Services (WRDS) Compustat database. It also features, despite a smaller total market capitalisation, more companies than the New York Stock Exchange, and is subject to considerably higher year-to-year volatility. The latter is especially interesting for analyses comparing annual differences in informational efficiency, which is why we opt for this exchange as a data source.

This provides us with a dataset featuring 809,195 entries for 4,905 companies, with associated company identifiers and dates. Missing values are a challenge commonly encountered in financial data, and have to be dealt with either through omission of affected entries or imputation methods. While the latter, despite their widespread use, are sometimes cautioned against, for example by \citet{Kofman2003}, the problem that we would encounter in our analysis is more fundamental:

We are, as Section~\ref{sec:overlapping_permutations_to_test_randomness} will detail in a bit, interested in the distribution of binary patterns, and the content of missing sections can be entirely unrelated to the pattern of missing entries, for example due to data collection issues stemming from technical difficulties limited to certain periods. As the question how these subtly changing binary patterns should be imputed is difficult to answer satisfactorily, we drop companies that feature missing monthly close prices within the time frame covered by the dataset. This leads to the omission of 417 companies, or approximately $8.50\%$, and is followed by a further cleaning step that drops companies that feature less than a year's worth of entries, resulting in another $5.86\%$ being sorted out, which is acceptable given that we investigate efficiency across the exchange and on the company level.

\begin{figure}[!tb]
\centering
\includegraphics[width=\columnwidth]{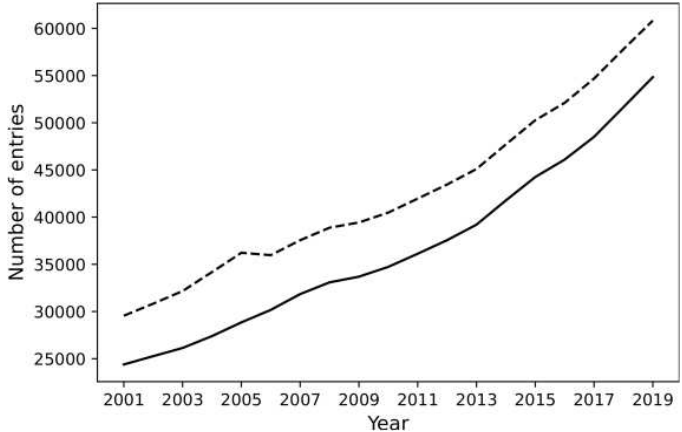}
\caption{Data points per calendar year. The figure shows entries for monthly stock prices available for Nasdaq-listed companies covering the years 2001--2019. The dashed line indicates the full dataset, whereas the solid line denotes the dataset with the omission of companies that feature missing values for price entries.}
\label{fig:figure_1}
\end{figure}

Figure~\ref{fig:figure_1} shows the number of entries per year, tracing the evolution of companies featured on the exchange over time, with dashed and solid lines indicating the dataset before and after the preprocessing, respectively. Aside from the increasing number of Nasdaq-listed public companies, two features stick out. The first is the slowing of growth around 2008, which can be explained by the impact of the Global Financial Crisis of 2007--2008 on IPOs \citep{Aktas2019}.

The second is the bump around 2005, which shows a slight decrease, and is mostly visible in the dataset before preprocessing. Natural explanations for this deviation include aftershocks of the Dotcom Bubble's burst a few years prior as well as the privatisation effect of the Sarbanes-Oxley Act in the United States following a series of corporate and accounting scandals, which regulates financial reporting and report keeping for public companies. This wave of privatisations for formerly public companies is demonstrated in the literature, for example by \citet{Engel2007}.

The effect on efficiency is still debated, as described in an overview by \citet{Bai2016}, although this is combined with a reported lack of empirical evidence for disclosure legislation leading to breaks in market informativeness. As our approach studies market efficiency regardless of contributing factors, this is not of direct concern, but our analysis shows an improvement in month-to-month efficiency for the year said act was passed, lasting until the Global Financial Crisis of 2007--2008.

\subsection{Data preprocessing and considerations}
\label{sec:data_preprocessing_and_considerations}

\begin{figure*}[!tb]
\centering
\includegraphics[width=0.825\textwidth]{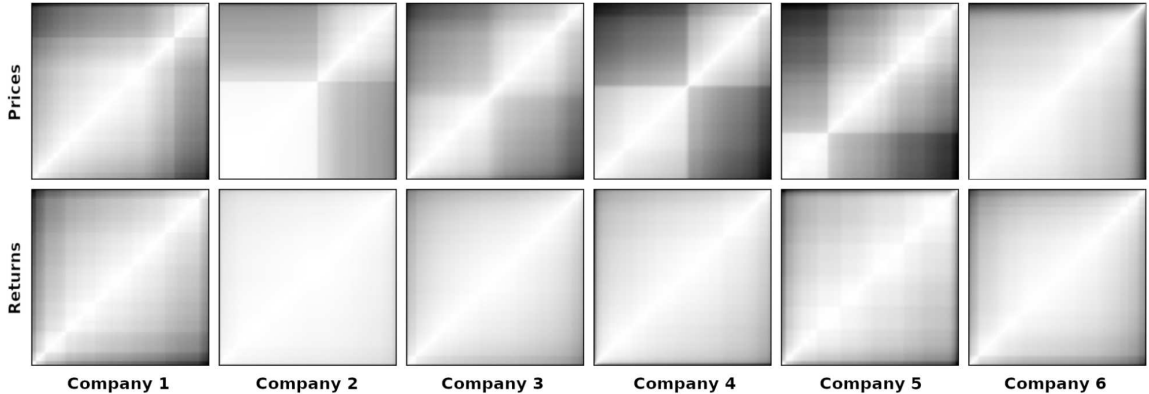}
\caption{Recurrence plots for stock prices and relative returns. Each column of the plot corresponds to one randomly sampled Nasdaq-listed company from a dataset covering the years 2001--2019. The first row shows recurrence plots for unprocessed stock prices, while the second row shows the same type of plot for logarithmic returns relative to the respective prior period's price for the same company.}
\label{fig:figure_2}
\end{figure*}

WRDS Compustat, as described in Section~\ref{sec:exchange_and_empirical_data_description}, provides both cumulative adjustment factors and total return factors for the purpose of price adjustment for any given time period, with the former being a ratio that enables per-share prices to be adjusted for all stock splits and dividends occurring subsequent to the end of a given period. Similarly, the latter represents a multiplication factor that includes cash-equivalent distributions along with reinvestment of dividends, as well as the compounding effect of dividends paid on reinvested dividends. Following the database's guidelines, we compute adjusted close prices from unadjusted prices $\hat{p}_{i, t}$, and for $\delta_{i, t}$ and $\gamma_{i, t}$ as the cumulative adjustment factor and the total return factor, respectively, as
\begin{equation}
p_{i, t} = \frac{\hat{p}_{i, t} \cdot \delta_{i, t}}{\gamma_{i, t}}.
\end{equation}
In the next step, we calculate the return by computing the natural logarithm of the price ratio between the current and prior period for given price series of length $N$,
\begin{equation}
r_{i, t} = \log_e \left( \frac{p_{i, t}}{p_{i, t - 1}} \right), \ \mathrm{with} \ t \in \{1, 2, \dots , N\}.
\end{equation}
Here, the logarithm takes the fact into account that individual stocks' price changes are partially dependent on price magnitudes \citep{Karpoff1987}. In order to visualise the relevance of working with returns instead of prices, Figure~\ref{fig:figure_2} shows recurrence plots for a random sample of companies from the dataset, with a recurrence plot $R_{n, m}$ for horizontal and vertical axes $n$ and $m$ generally being calculated as
\begin{equation}
R_{n, m} = \Theta ( \epsilon - || \overrightarrow{v}_n - \overrightarrow{v}_m || ),
\end{equation}
where $\overrightarrow{v}$ is a phase space trajectory, $\epsilon$ is a binarisation threshold, and $\Theta$ is the Heaviside step function. Recurrence plots are frequently used in both statistics and chaos theory to image the periodic nature of phase space trajectories, meaning similar areas being visited in such a space. More informally and in our case, it shows the distance between points along a time series, omitting the binarisation and visualising recurrences as times a trajectory returns to a previous value.

The first row of the figure, corresponding to a trajectory $\overrightarrow{p}_{i}$, shows slightly darkened upper-left and lower-right corners in the first row indicating slow adiabatic changes in line with the framing of market efficiency as a random walk with drift by \citet{Fama1970}. As we can see, raw prices are less than ideal in terms of their homogeneity, and the calculation of logarithmic returns in the second row generally alleviates these problems. In the final step, we binarise the return series using the median $\check{r}_{t}$,
\begin{equation}
b_{i, t} = \left\{
        \begin{array}{ll}
            1 & \quad \mathrm{if} \ r_{i, t} > \check{r}_{i} \ , \\
            0 & \quad \mathrm{else}.
        \end{array}
    \right.
\end{equation}
The choice of the median over the arithmetic mean follows \citet{Doyle2013}, as this option yields equal numbers of ones and zeroes in the resulting binary array, with an offset of one for uneven lengths. This binarisation also takes care of heteroskedasticity as the lack of variance homogeneity along the evolution of a given time series, which corresponds to return volatility in our case. The presence of heteroskedasticity in markets is well-known in both the financial and operational research literature \citep{Mandelbrot1967,  Lamoureux1990, Fabozzi2017, Meligkotsidou2019}. 

Lastly, another well-known effect in financial time series is momentum, meaning positive autocorrelation, which is generally described as a premier anomaly of the EMH \citep{Fama1970}. Financial research sometimes uses ``runs tests'' to check for unbroken sequences of upward or downward movements, which struggle to find oscillations between positive and negative autocorrelations, as they cancel each other out. The overlapping permutations test described in the following Section~\ref{sec:overlapping_permutations_to_test_randomness} is not subject to this shortcoming, and also able to check for patterns beyond such unbroken sequences.

\subsection{Overlapping permutations to test randomness}
\label{sec:overlapping_permutations_to_test_randomness}

Originally developed as the generalised serial test by \citet{Good1967}\footnote{As an interesting anecdote, Irving J. Good served alongside Alan Turing at Bletchley Park during WWII, and later also was a consultant for Stanley Kubrick's science fiction classic ``2001: A Space Odyssey''.}, this approach is based on a small number of earlier works considering the use of permutations to assess randomness \citep{Kendall1938, Bartlett1951, Good1953}. It tests for the equiprobability of $k^\nu$ separate $\nu$-nomes, or permutations of length $\nu$, for $k$ possible letters, and later found entry into the Diehard Battery of Tests for RNGs, here it is still one of the recommended core constituents of today's testing suites for pseudo-RNGs \citep{Marsaglia2002, Luengo2021}. In its binary variation, we set $k = 2$ and calculate, in an analogy to $\chi^2$ for multinomial distributions,
\begin{equation}
\begin{split}
\psi^2_\nu &= \sum_{i=1}^{2^\nu} \frac{(n_i - \lambda)^2}{\lambda}, \ \mathrm{with} \\
\lambda &= \frac{N - \nu + 1}{k}
\end{split}
\label{eq:psi_square}
\end{equation}
as the expectation for the frequency of each unique pattern under the assumption of uniformity. As $\Psi_\nu^2$ does not have an asymptotic tabular $\chi^2$ distribution due to a violation of the assumption of independence caused by the overlap of windows, \citet{Good1967} propose first differences to alleviate this problem as
\begin{equation}
\nabla \psi_\nu^2 = \psi_\nu^2 - \psi_{\nu - 1}^2.
\end{equation}
This statistic fulfils an asymptotic tabular $\chi^2$ distribution; and taking the second difference also fulfils asymptotic independence, meaning that we can calculate
\begin{equation}
\begin{split}
\nabla^2 \psi_\nu^2 &= \nabla \psi_\nu^2 - \nabla \psi_{\nu - 1}^2 \\
&= \psi_\nu^2 - 2 \psi_{\nu - 1}^2 + \psi_{\nu - 2}^2,
\end{split}
\label{eq:second_difference}
\end{equation}
with $\xi_\nu = \nabla^2 2^\nu = 2^\nu - 2 \cdot 2^{\nu - 1} + 2^{\nu - 2} = 2^{\nu - 2}$ as the associated degrees of freedom for $\nu \geq 2$. While the difference of one in separate $\psi_\nu^2$ values in the case of uneven array lengths from Section~\ref{sec:data_preprocessing_and_considerations} gives rise to the question of the impact, most terms cancel out in sums over $\nabla^2 \psi_\nu^2$, as shown by \citet{Doyle2013}. Following the prior literature, we make use of $\nu \in \{1, 2, \dots , 8\}$ in our tests, and rely on the second differences due to its improved suitability for testing for uniform randomness \citep{Marsaglia2005}. 

\section{Empirical experiments and results}
\label{sec:empirical_experiments_and_results}

\subsection{Tests of monthly information incorporation}
\label{sec:tests_of_monthly_information_incorporation}

In the first step, we calculate $\psi_\nu^2$ values for our dataset, split into two experimental streams for company-separated and year-separated arrays, respectively. As we are dealing with a set of 4,225 company-associated stocks, summary statistics are computed for these measurements. The upper part of Table~\ref{tab:table_1} shows these results on a company-separated level, listing the arithmetic mean and standard deviation for window sizes $\nu \in \{1, 2, \dots , 8\}$. In addition, as large-valued subsets will become relevant further down, the table also shows the respective maximum per window size. 

Results for $\nu = 1$ correspond to the measurement of single binary values, and are expectedly slightly larger than for subsequent year-level experiments due to the smaller average array length for monthly entries over the investigated time frame.

\begin{table*}[htb!]
\begin{small}
\setlength{\tabcolsep}{1.15pc}
\caption{Psi-square statistic per window size. The table shows, for window sizes $\nu \in \{1, 2, \dots, 8\}$, the mean, standard deviation, and maximum for $\psi^2$ values for monthly data of Nasdaq-listed companies in the 2001--2019 time frame. The upper and lower parts show results for data separated by company and year, respectively, as well as $\psi^2$ values for each year.\textsuperscript{a}}
{\begin{tabular}{lrrrrrrrr} \toprule
$\nu$ & 1 & 2 & 3 & 4 & 5 & 6 & 7 & 8 \\ \midrule
$\overline{\psi_\nu^2}$ & 0.16 & 1.63 & 5.24 & 13.15 & 29.28 & 61.84 & 127.64 & 258.64 \\
$\sigma\left({\psi_\nu^2}\right)$ & 1.72 & 4.28 & 8.25 & 14.01 & 22.71 & 35.94 & 57.74 & 93.19 \\
$\max(\psi_\nu^2)$ & 47.51 & 107.48 & 222.87 & 393.09 & 694.66 & 1218.16 & 2112.62 & 3606.53 \\
\midrule
2001 & 0 & 3.19 & 62.97 & 179.27 & 337.94 & 533.66 & 982.93 & 1562.51 \\
2002 & 3.96 $\cdot 10^{-5}$ & 14.73 & 33.69 & 70.20 & 173.48 & 332.81 & 646.53 & 1044.54 \\
2003 & 3.44 $\cdot 10^{-4}$ & 1.59 & 9.24 & 32.07 & 72.71 & 168.11 & 392.00 & 750.94 \\
2004 & 0 & 0.29 & 8.35 & 18.12 & 83.48 & 165.97 & 291.73 & 550.74 \\
2005 & 0 & 0.75 & 4.49 & 80.19 & 167.02 & 304.36 & 512.86 & 841.06 \\
2006 & 2.98 $\cdot 10^{-4}$ & 0.35 & 3.819 & 20.92 & 60.21 & 130.61 & 262.12 & 474.63 \\
2007 & 3.14 $\cdot 10^{-5}$ & 6.82 & 39.67 & 78.64 & 130.26 & 217.29 & 354.00 & 597.86 \\
2008 & 0 & 18.81 & 174.88 & 375.87 & 812.36 & 1409.31 & 2602.95 & 4038.67 \\
2009 & 0 & 39.60 & 202.52 & 406.63 & 979.43 & 1704.42 & 2647.44 & 3786.23 \\
2010 & 0 & 28.05 & 58.63 & 482.49 & 1368.27 & 2449.59 & 3854.00 & 5694.62 \\
2011 & 0 & 28.19 & 110.34 & 231.57 & 394.20 & 743.01 & 1193.87 & 1853.62 \\
2012 & 0 & 14.62 & 44.44 & 83.91 & 215.15 & 374.13 & 628.24 & 952.90 \\
2013 & 2.55 $\cdot 10^{-5}$ & 48.44 & 180.98 & 363.20 & 658.10 & 991.05 & 1471.73 & 2054.90 \\
2014 & 2.39 $\cdot 10^{-5}$ & 248.53 & 594.81 & 973.97 & 1413.72 & 1928.59 & 2607.41 & 3472.00 \\
2015 & 2.26 $\cdot 10^{-5}$ & 1.04 & 29.96 & 100.75 & 248.79 & 457.51 & 939.28 & 1737.03 \\
2016 & 0 & 33.84 & 126.71 & 229.33 & 547.28 & 971.32 & 1513.02 & 2172.17 \\
2017 & 0 & 7.75 & 29.92 & 82.13 & 167.52 & 342.50 & 575.70 & 917.13 \\
2018 & 1.94 $\cdot 10^{-5}$ & 0.27 & 350.79 & 728.50 & 1144.63 & 1875.68 & 2831.41 & 4152.08 \\
2019 & 0 & 140.65 & 316.36 & 556.10 & 846.33 & 1291.62 & 1930.10 & 2693.00 \\\\
$\overline{\psi_\nu^2}$ & 4.20 $\cdot 10^{-5}$ & 33.55 & 125.40 & 268.10 & 516.89 & 862.71 & 1380.91 & 2070.87 \\
$\sigma\left({\psi_\nu^2}\right)$ & 9.70 $\cdot 10^{-5}$ & 59.70 & 149.53 & 259.14 & 439.46 & 696.19 & 1039.12 & 1468.50 \\
$\max(\psi_\nu^2)$ & 3.44 $\cdot 10^{-4}$ & 248.53 & 594.81 & 973.97 & 1413.72 & 2449.59 & 3853.99 & 5694.62 \\
\bottomrule
\end{tabular}}

\footnotesize{\textsuperscript{a}The calculation of $\psi^2_\nu$ follows Equation~\ref{eq:psi_square}.}
\label{tab:table_1}
\end{small}
\end{table*}

\begin{figure}[!tb]
\centering
\includegraphics[width=\columnwidth]{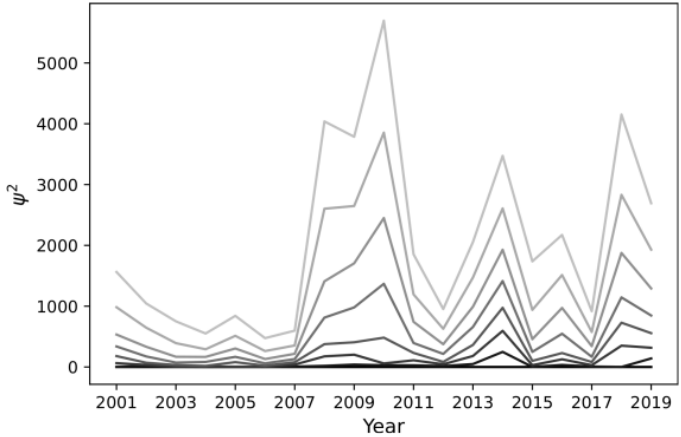}
\caption{Psi-square statistic per calendar year. The figure shows the evolution of $\psi^2$ values for Nasdaq-listed companies in the 2001--2019 time frame, with shifting window sizes $\nu \in \{1, 2, \dots, 8\}$. The statistic follows Equation~\ref{eq:psi_square}, with lighter shades of grey corresponding to higher values for the window size.}
\label{fig:figure_3}
\end{figure}

Next, we repeat the same experiment with entries separated by their year, leading to the results listed in the lower part of Table~\ref{tab:table_1}. As opposed to the company-separated case, 19 entries easily lend themselves to being listed individually in a table, in addition to the summary statistics already used before. The $\psi_\nu^2$ values paint a very diverse picture in terms of the year-to-year volatility of measured pattern retrieval, with both higher means and maxima. In part, this can be explained by the possibility of patterns occurring within the constraints of a given annual time frame, whereas company-separated measurements spanning the entire time period of the dataset offer an avenue to even these pattern distributions out. 

While the means in the upper part of Table~\ref{tab:table_1} closely trace the values reported by \citet{Doyle2013} for the Nasdaq Composite index, validating both the implementation and the stability of these analyses over different time frames, year-to-year analyses paint a different picture. Given these findings, the question arises whether different window sizes correlate in terms of these results. 

\begin{figure}[!tb]
\centering
\includegraphics[width=0.825\columnwidth]{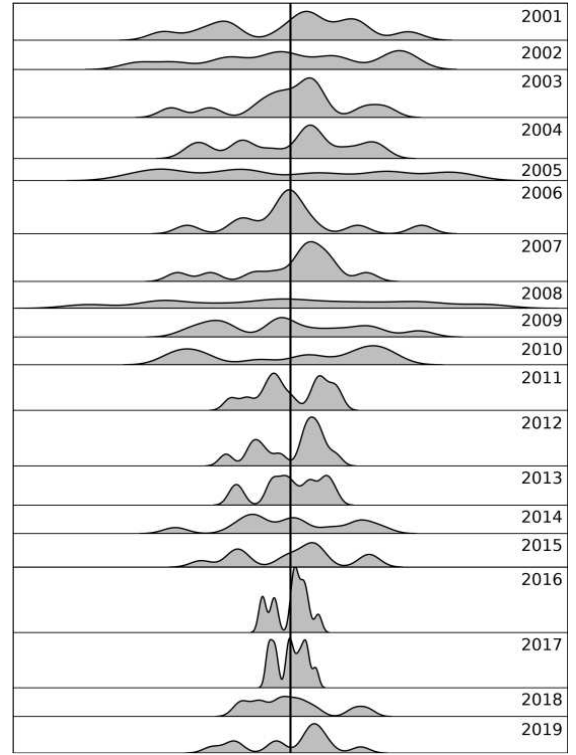}
\caption{Kernel density estimates of monthly variability. The figure shows the distribution of binarised monthly returns per year, reshaped into month-wise columns and summed over each column.}
\label{fig:figure_4}
\end{figure}

We plot $\psi_\nu^2$ values for window sizes $\nu \in \{ 1, 2, \dots , 8 \}$ in Figure~\ref{fig:figure_3} to see how indicators for the presence of recurring patterns relate to different pattern lengths. The plot demonstrates mostly strong co-movements across window sizes, steering us towards the notion that when patterns of one size are recurring above uniformity, so are those corresponding to other window sizes.

In the next step, we calculate second differences as per Equation~\ref{eq:second_difference}, again for both company-separated and year-separated datasets. Given the mathematical guarantees of $\nabla^2 \psi^2_\nu$ values outlined in Section~\ref{sec:overlapping_permutations_to_test_randomness} and further described by \citet{Good1967}, we can now make use of critical $\chi_2$ values for given degrees of freedom at the commonly employed 5\% level for statistical significance. Second differences that fail to meet this threshold, thus not supporting the discarding of the null hypothesis of uniform randomness, are indicated in bold.

The upper part of Table~\ref{tab:table_2} shows results for company-separated data. Interestingly, while the arithmetic means indicate uniform randomness, the combined $\chi_2$ values for multiplied degrees of freedom of $|A| \times \xi_\nu$, for lengths of given arrays A, do not share this result. This leads us to the suspicion that there are small subsets of highly inefficient stocks that skew the sums upwards, which is further supported by the calculation of proportions of statistically significant measures in the table.

\begin{table*}[htb!]
\begin{small}
\setlength{\tabcolsep}{1.15pc}
\caption{Second difference for increasing degrees of freedom. The table shows, for degrees of freedom $\xi \in \{2, 4, 8, 16, 32, 64\}$, the mean and standard deviation for $\nabla^2 \psi^2_\nu$ values for monthly data of Nasdaq-listed companies in the 2001--2019 time frame, as well as the combined $\chi^2$ statistic. The upper part shows results for data separated by company, as well as combined $\chi^2$ statistics for percentual omissions of the highest contributors. The lower part shows results for data separated by year, as well as $\nabla^2 \psi^2_\nu$ values for each year. Results failing the threshold for significance at the 5\% level are indicated in bold.\textsuperscript{a}}
{\begin{tabular}{lrrrrrr} \toprule
 & $\nabla^2 \psi_3^2$ & $\nabla^2 \psi_4^2$ & $\nabla^2 \psi_5^2$ & $\nabla^2 \psi_6^2$ & $\nabla^2 \psi_7^2$ & $\nabla^2 \psi_8^2$ \\ 
 & $\xi = 2$ & $\xi = 4$ & $\xi = 8$ & $\xi = 16$ & $\xi = 32$ & $\xi = 64$ \\ \midrule
$\overline{\nabla^2 \psi^2_\nu}_{\mathrm{, firms}}$ & \textbf{2.15} & \textbf{4.30} & \textbf{8.21} & \textbf{16.45} & \textbf{33.11} & \textbf{65.44} \\
$\sigma \left(\nabla^2 \psi^2_\nu\right)_{\mathrm{firms}}$ & 2.49 & 3.39 & 5.03 & 7.62 & 11.89 & 17.93 \\\\
$\sum \chi^2$ & 9067.27 & 18188.26 & 34676.73 & 69480.95 & 139904.27 & 276477.57 \\
$\sum \chi^2_{- 1\%}$ & \textbf{8403.91} & 17301.45 & \textbf{33297.16} & \textbf{67200.00} & 136140.16 & 270135.37 \\
$\sum \chi^2_{- 2\%}$ & \textbf{8015.03} & \textbf{16729.07} & \textbf{32454.35} & \textbf{65791.98} & 133849.35 & \textbf{265978.62} \\
$\sum \chi^2_{- 3\%}$ & \textbf{7678.14} & \textbf{16210.66} & \textbf{31670.69} & \textbf{64460.44} & \textbf{131629.57} & \textbf{261950.75} \\
$| \chi^2_{p < 0.05} | \ / \ | \chi^2 |$ & 5.68 $\cdot 10^{-2}$ & 6.13 $\cdot 10^{-2}$ & 5.21 $\cdot 10^{-2}$ & 6.04 $\cdot 10^{-2}$ & 6.20 $\cdot 10^{-2}$ & 6.93 $\cdot 10^{-2}$ \\ 
\midrule
2001 & 56.58 & 56.53 & 42.37 & 37.05 & 253.54 & 130.32 \\
2002 & \textbf{4.24} & 17.53 & 66.78 & 56.04 & 154.39 & 84.29 \\
2003 & 6.05 & 15.19 & 17.80 & 54.77 & 128.49 & 135.05 \\
2004 & 7.78 & \textbf{1.70} & 55.59 & \textbf{17.14} & \textbf{43.26} & 119.69 \\
2005 & \textbf{3.00} & 71.95 & \textbf{11.13} & 50.51 & 71.16 & \textbf{81.00} \\
2006 & \textbf{3.13} & 13.63 & 22.19 & 31.11 & 61.11 & 107.16 \\
2007 & 6.03 & \textbf{6.12} &  \textbf{12.66} & 35.40 & 49.67 & 242.07 \\
2008 & 137.25 & 44.92 & 235.50 & 160.46 & 596.71 & 195.76 \\
2009 & 123.32 & 41.18 & 368.70 & 152.18 & 218.04 & 436.23 \\
2010 & \textbf{2.54} & 393.27 & 461.92 & 195.54 & 323.08 & 208.90 \\
2011 & 53.97 & 39.07 & 41.40 & 186.18 & 102.04 & \textbf{70.55} \\
2012 & 15.20 & 9.65 & 91.77 & 27.74 & 95.13 & 102.49 \\
2013 & 84.11 & 49.68 & 112.68 & 38.05 & 147.74 & 185.78 \\
2014 & 97.75 & 32.88 & 60.58 & 75.14 & 163.94 & 315.98 \\
2015 & 27.89 & 41.85 & 77.26 & 60.68 & 273.04 & 117.46 \\
2016 & 59.02 & 9.75 & 215.34 & 106.08 & 117.66 & 108.23 \\
2017 & 14.42 & 30.04 & 33.19 & 89.58 & 58.23 & 364.93 \\
2018 & 350.25 & 27.19 & 38.43 & 314.91 & 224.69 & 364.93 \\
2019 & 35.06 & 64.03 & 50.49 & 155.06 & 193.18 & 124.43 \\\\
$\overline{\nabla^2 \psi^2_\nu}_{\mathrm{, years}}$ & 58.29 & 50.85 & 106.09 & 97.03 & 172.37 & 171.77 \\
$\sigma \left(\nabla^2 \psi^2_\nu\right)_{\mathrm{years}}$ & 80.01 & 83.08 & 122.54 & 75.93 & 127.52 & 99.14 \\\\
$\sum \chi^2$ & 1107.59 & 966.17 & 2015.79 & 1843.62 & 3275.09 & 3263.58 \\
$| \chi^2_{p < 0.05} | \ / \ | \chi^2 |$ & 0.79 & 0.89 & 0.89 & 0.95 & 0.95 & 0.89 \\ \bottomrule
\end{tabular}}

\footnotesize{\textsuperscript{a}The calculation of $\nabla^2 \psi_\nu^2$ follows Equation~\ref{eq:second_difference}.}
\label{tab:table_2}
\end{small}
\end{table*}

We confirm this effect by dropping sufficiently small percentages of the highest contributors and reevaluating the measures for combined $\chi_2$ values. When doing so, it is important to readjust critical $\chi^2$ values to account for the slightly reduced array sizes. The fourth to penultimate rows in the upper part of Table~\ref{tab:table_2} demonstrate how small percentages (1\%, 2\%, 3\%) result in statistical insignificance for the combined $\chi_2$ values for different window sizes, starting with $\nu = 3, 5, 6$ for 1\%. We extend the coverage by $\nu = 4, 8$ for 2\%, and finally add $\nu = 7$ for 3\%, bringing the sums in line with the arithmetic means in terms of their support for uniform randomness.

Next, in the lower part of Table~\ref{tab:table_2}, we repeat the same measures for the year-separated dataset as for $\psi_\nu^2$ values before in the lower part of Table~\ref{tab:table_1}. We see, just like for the company-separated results, statistically significant deviations from uniform randomness in the arithmetic means, which we have now established to be due to small subsets of companies with pattern-heavy stock behaviour. This results, when viewing measures at the annual level, in a high proportion of significant results, although these statistics vary starkly between years and demonstrate annual variations in market efficiency as measured through recurring patterns in this paper.

Finally, as a complementary visualisation, we transform each year's array into its constituent months to separate stocks, and plot kernel density estimates of the resulting array sums in Figure~\ref{fig:figure_4}. This translates to each column-wise sum being a count of ones per month and year, where uniformly-random distributions would approximate a narrow distribution around the mean. The horizontal axis is scaled based on array lengths to maintain comparability, and distributions are centred around the mean indicated by a solid vertical line. While not a perfect approximation by any means, the evolution of the count spreads roughly follows the time series in Figure~\ref{fig:figure_3}, including a broad distribution corresponding to increased intra-market cross-correlations during the Global Financial Crisis of 2007--2008 \citep{Zheng2012}.

\subsection{Comparison to measurements of daily data}
\label{sec:comparison_to_measurements_of_daily_data}

In contrast to \citet{Doyle2013} and \citet{Noakes2016}, who perform analyses on 76 and 111 time series, respectively, our analysis covers the entire exchange and operate on 4,905 instruments as described in Section~\ref{sec:exchange_and_empirical_data_description}. While this makes the use of monthly instead of daily close prices a natural choice, the computational expense to repeat our analysis for daily time steps, which brings the number of entries from 809,195 to 18,832,546, is beneficial as a comparison. Consequently, we obtain daily close prices for the same stocks and years, and repeat our experiments to test for statistically significant deviations from uniform randomness.

Table~\ref{tab:table_3} shows the results of this additional analysis. Measurements for the arithmetic means and standard deviations for company-separated data stay almost identical to both our monthly results and \citet{Doyle2013}'s values for the Nasdaq Composite, demonstrating the broad consistency of our results for different time frames. One interesting difference to our previous results is the dropping of the highest 5\% of inefficient contributors to achieve statistically significant measurements of efficiency across all degrees of freedom. 

\begin{table*}[htb!]
\begin{small}
\setlength{\tabcolsep}{1.15pc}
\caption{Second difference for increasing degrees of freedom. The table shows, for degrees of freedom $\xi \in \{2, 4, 8, 16, 32, 64\}$, the mean and standard deviation for $\nabla^2 \psi^2_\nu$ values for daily data of Nasdaq-listed companies in the 2001--2019 time frame, as well as the combined $\chi^2$ statistic. The upper part shows results for data separated by company, as well as combined $\chi^2$ statistics for percentual omissions of the highest contributors. The lower part shows results for data separated by year, as well as $\nabla^2 \psi^2_\nu$ values for each year. Results failing the threshold for significance at the 5\% level are indicated in bold.\textsuperscript{a}}
{\begin{tabular}{lrrrrrr} \toprule
 & $\nabla^2 \psi_3^2$ & $\nabla^2 \psi_4^2$ & $\nabla^2 \psi_5^2$ & $\nabla^2 \psi_6^2$ & $\nabla^2 \psi_7^2$ & $\nabla^2 \psi_8^2$ \\ 
 & $\xi = 2$ & $\xi = 4$ & $\xi = 8$ & $\xi = 16$ & $\xi = 32$ & $\xi = 64$ \\ \midrule
$\overline{\nabla^2 \psi^2_\nu}_{\mathrm{, firms}}$ & \textbf{2.44} & \textbf{4.48} & \textbf{8.53} & \textbf{16.56} & \textbf{32.75} & \textbf{64.88} \\
$\sigma \left(\nabla^2 \psi^2_\nu\right)_{\mathrm{firms}}$ & 2.56 & 3.44 & 4.61 & 6.52 & 9.71 & 14.53 \\\\
$\sum \chi^2$ & 10396.00 & 19075.23 & 36296.73 & 70430.19 & 139315.84 & 276003.70 \\
$\sum \chi^2_{- 1\%}$ & 9756.11 & 18199.13 & 35141.65 & 68662.60 & 136419.66 & 271053.15 \\
$\sum \chi^2_{- 2\%}$ & 9306.81 & 17581.32 & 34304.87 & 67314.86 & \textbf{134148.97} & \textbf{267121.43} \\
$\sum \chi^2_{- 3\%}$ & 8901.99 & 17016.34 & 33501.13 & \textbf{66023.67} & \textbf{131939.71} & \textbf{263231.51} \\
$\sum \chi^2_{- 4\%}$ & 8547.31 & \textbf{16512.25} & \textbf{32758.54} & \textbf{64816.08} & \textbf{129856.88} & \textbf{59525.77} \\
$\sum \chi^2_{- 5\%}$ & \textbf{8210.39} & \textbf{16032.74} & \textbf{32033.78} & \textbf{63628.77} & \textbf{127782.30} & \textbf{255810.71} \\
$| \chi^2_{p < 0.05} | \ / \ | \chi^2 |$ & 8.74 $\cdot 10^{-2}$ & 7.97 $\cdot 10^{-2}$ & 7.05 $\cdot 10^{-2}$ & 6.46 $\cdot 10^{-2}$ & 6.42 $\cdot 10^{-2}$ &6.54 $\cdot 10^{-2}$ \\ 
\midrule
2001 & 66.44 & 51.39 & 104.54 & 66.91 & 204.98 & 207.94 \\
2002 & 51.36 & 62.46 & 103.91 & 157.59 & 193.32 & 244.34 \\
2003 & \textbf{1.51} & 54.27 & 99.86 & 85.91 & 156.24 & 314.02 \\
2004 & \textbf{0.028} & 41.98 & 86.17 & 109.40 & 64.67 & 179.81 \\
2005 & 22.51 & 124.63 & 156.29 & 86.30 & 78.70 & 217.40 \\
2006 & 26.29 & 76.96 & 21.28 & 86.30 & 205.94 & 189.65 \\
2007 & 38.36 & 81.70 & 352.04 & 356.78 & 108.37 & 393.37 \\
2008 & 51.40 & 68.13 & 126.97 & 531.60 & 377.72 & 1130.57 \\
2009 & 160.50 & 31.54 & 202.79 & 557.00 & 280.62 & 730.46 \\
2010 & 82.33 & \textbf{9.00} & 131.26 & 267.02 & 633.03 & 1108.18 \\
2011 & 33.55 & 743.2 & 310.15 & 1555.89 & 1163.54 & 1518.06 \\
2012 & 17.84 & 10.15 & 440.16 & 204.99 & 258.52 & 275.35 \\
2013 & 69.53 & 101.75 & 87.91 & 121.29 & 373.04 & 387.18 \\
2014 & 44.02 & 112.24 & 61.78 & 88.16 & 215.85 & 335.25 \\
2015 & 221.47 & 360.49 & 280.11 & 390.58 & 282.27 & 538.77 \\
2016 & 136.87 & 156.62 & 176.59 & 157.88 & 561.84 & 528.90 \\
2017 & 18.66 & 71.25 & 68.46 & 154.29 & 125.00 & 218.86 \\
2018 & 79.27 & 282.88 & 174.81 & 163.81 & 221.11 & 205.80 \\
2019 & 17.78 & 172.42 & 225.25 & 149.04 & 254.40 & 344.64 \\\\
$\overline{\nabla^2 \psi^2_\nu}_{\mathrm{, years}}$ & 70.51 & 137.53 & 168.96 & 281.82 & 303.11 & 477.29 \\
$\sigma \left(\nabla^2 \psi^2_\nu\right)_{\mathrm{years}}$ & 66.45 & 167.10 & 106.89 & 332.19 & 248.76 & 370.28 \\\\
$\sum \chi^2$ & 1339.71 & 2613.12 & 3210.33 & 5354.63 & 5759.14 & 9068.57 \\
$| \chi^2_{p < 0.05} | \ / \ | \chi^2 |$ & 0.89 & 0.95 & 1.00 & 1.00 & 1.00 & 1.00 \\ \bottomrule
\end{tabular}}

\footnotesize{\textsuperscript{a}The calculation of $\nabla^2 \psi_\nu^2$ follows Equation~\ref{eq:second_difference}.}
\label{tab:table_3}
\end{small}
\end{table*}

Given the smaller time steps, more stocks have a chance to contribute highly inefficient periods to the overall result, which is confirmed by the listed higher shares of inefficient contributors for degrees of freedom requiring additional percentages to be dropped. While not unexpected, this provides a useful insight into the slight differences that data granularity can have on analyses.

One difference of particular interest is the clear concentration of inefficiency for higher degrees versus lower degrees of freedom, which indicates shorter non-random patterns being more spread out across instruments for daily data. Similarly, the results for the year-separated dataset repeat the previously observed deviations from uniform randomness in the arithmetic means, and the stronger presence of an inefficient subset leads to a relative elevation in year-to-year measurements.

Taking into account the known differences in market efficiency in favour of longer time scales \citep[see, for example,][]{Kim2008, Rodriguez2014}, monthly and daily close prices reasonably mirror each other in terms of the overall findings and implications. This first application of the approach used in our experiments to varying data frequencies encourages the analysis of different time frames in related works, which we touch upon in Section~\ref{sec:discussion}.

\subsection{Comparison to pseudo-random numbers}
\label{sec:comparison_to_results_from_chaotic_maps}

\begin{table*}[htb!]
\begin{small}
\setlength{\tabcolsep}{1.15pc}
\caption{Psi-square statistic per window size. The table shows, for window sizes $\nu \in \{1, 2, \dots, 8\}$, the mean, standard deviation, and maximum of $\psi^2_\nu$ for generated pseudo-random numbers. Rows 1--3 and rows 4--6 cover a set modelled on the year-separated and firm-separated dataset, respectively.\textsuperscript{a}}
{\begin{tabular}{lrrrrrrrr} \toprule
$\nu$ & 1 & 2 & 3 & 4 & 5 & 6 & 7 & 8 \\ \midrule
$\overline{\psi_\nu^2}_\mathrm{,years}$ & 1.23 $\cdot 10^{-5}$ & 0.60 &  2.95 & 9.84 & 23.38 & 51.45 & 110.37 & 231.18 \\
$\sigma\left({\psi_\nu^2}\right)_\mathrm{years}$ & 1.51 $\cdot 10^{-5}$ & 0.54 & 2.63 & 6.35 & 11.56 & 19.77 & 30.75 & 45.76 \\
$\max(\psi_\nu^2)_\mathrm{years}$ & 210.51 & 225.86 & 218.95 & 244.50 & 201.04 & 210.43 & 184.10 & 201.31 \\\\
$\overline{\psi_\nu^2}_\mathrm{,firms}$ & 6.93 $\cdot 10^{-3}$ & 1.07 & 4.17 & 11.32 & 26.37 & 57.47 & 120.64 & 248.09 \\
$\sigma\left({\psi_\nu^2}\right)_\mathrm{firms}$ & 1.13 $\cdot 10^{-2}$ & 1.47 & 3.51 & 6.50 & 10.81 & 16.67 & 25.08 & 37.02 \\
$\max(\psi_\nu^2)_\mathrm{firms}$ & 0.08 & 13.89 & 31.00 & 65.27 & 139.00 & 274.40 & 513.21 & 949.11 \\
\bottomrule
\end{tabular}}

\footnotesize{\textsuperscript{a}The calculation of $\psi^2_\nu$ follows Equation~\ref{eq:psi_square}.}
\label{tab:table_4}
\end{small}
\end{table*}

\begin{table*}[htb!]
\begin{small}
\setlength{\tabcolsep}{1.15pc}
\caption{Results for second differences for increasing degrees of freedom. The table shows, for degrees of freedom $\xi \in \{2, 4, 8, 16, 32, 64\}$, summary statistics for $\nabla^2 \psi^2_\nu$ values across generated pseudo-random numbers. Rows 1--4 and rows 5--8 cover a set modelled on the year-separated and firm-separated dataset, respectively. Results failing the threshold for significance at the 5\% level are marked in bold.\textsuperscript{a}}
{\begin{tabular}{lrrrrrr} \toprule
& $\nabla^2 \psi_3^2$ & $\nabla^2 \psi_4^2$ & $\nabla^2 \psi_5^2$ & $\nabla^2 \psi_6^2$ & $\nabla^2 \psi_7^2$ & $\nabla^2 \psi_8^2$ \\ 
& $\xi = 2$ & $\xi = 4$ & $\xi = 8$ & $\xi = 16$ & $\xi = 32$ & $\xi = 64$ \\ \midrule
$\overline{\nabla^2 \psi^2_\nu}_\mathrm{,years}$ & \textbf{1.76} & \textbf{4.52} & \textbf{6.66} & \textbf{14.52} & \textbf{30.85} & \textbf{61.90} \\
$\sigma \left(\nabla^2 \psi^2_\nu\right)_\mathrm{years}$ & 2.42 & 3.30 & 3.01 & 6.15 & 7.02 & 9.83 \\\\
$\sum \chi^2_\mathrm{years}$ & \textbf{33.38} & \textbf{85.96} & \textbf{126.62} & \textbf{275.87} & \textbf{586.19} & \textbf{1176.06} \\
$| \chi^2_{p < 0.05} | \ / \ | \chi^2 |$ & 5.26 $\cdot 10^{-2}$ & 10.53 $\cdot 10^{-2}$ & 0 & 5.26 $\cdot 10^{-2}$ & 0 & 0 \\\\
$\overline{\nabla^2 \psi^2_\nu}_\mathrm{,firms}$ & \textbf{2.04} & \textbf{4.04} & \textbf{7.91} & \textbf{16.05} & \textbf{32.07} & \textbf{64.28} \\
$\sigma \left(\nabla^2 \psi^2_\nu\right)_\mathrm{firms}$ & 2.02 & 2.79 & 4.09 & 5.68 & 7.96 & 11.2 \\\\
$\sum \chi^2_\mathrm{firms}$ & \textbf{8621.46} & \textbf{17066.77} & \textbf{33402.40} & \textbf{67826.51} & \textbf{135477.96} & \textbf{271599.01} \\
$| \chi^2_{p < 0.05} | \ / \ | \chi^2 |$ & 5.04 $\cdot 10^{-2}$ & 4.83 $\cdot 10^{-2}$ & 4.88 $\cdot 10^{-2}$ & 4.85 $\cdot 10^{-2}$ & 4.31 $\cdot 10^{-2}$ & 4.78 $\cdot 10^{-2}$ \\ \bottomrule
\end{tabular}}

\footnotesize{\textsuperscript{a}The calculation of $\nabla^2 \psi_\nu^2$ follows Equation~\ref{eq:second_difference}.}
\label{tab:table_5}
\end{small}
\end{table*}

When assessing the findings of the previous sections, a natural question is that of measurements one would expect from a uniformly-random distribution. This allows for a direct comparison to numerical results for our methodology that represent the case of the null hypothesis, and well-established pseudo-RNGs can be used as a baseline in our study's context of markets as an RNG analogue.

In terms of broader applications in programming languages, the MT19937 implementation of the Mersenne Twister algorithm has long been the a standard pseudo-RNG since its original inception as an answer to then-current flaws in older generators \citep{Matsumoto1998}. In more recent years, however, other general-purpose algorithms have been developed and begun to supplant its reign.

One example is the family of permuted congruential generators (PCG) introduced by \citet{Oneill2014}. The PCG64 implementation found widespread adoption, and was made the default generator used by the NumPy mathematical library as of version 1.17 in 2019. Among the reasons for this adoption are the passing of the TestU01 suite with zero failures, which distinguishes it from the Mersenne Twister algorithm as the prior default \citep{LEcuyer2007}.

Using the PCG64 implementation to repeat our experiments from Section~\ref{sec:tests_of_monthly_information_incorporation}, Table~\ref{tab:table_4} shows that the means for $\psi^2_\nu$ are very close to those for firm-separated values in the upper part of Table~\ref{tab:table_1}, with slightly higher standard deviations, while both means and standard deviations are notably lower for the year-separated dataset. In both cases, the respective maxima are considerably lower than those for the empirical Nasdaq data, underlining the previously noted impact of inefficient subsets.

For second differences $\nabla^2 \psi^2_\nu$, Table~\ref{tab:table_5} shows the same statistical metrics as before, for firm-separated values, as the lower part of Table~\ref{tab:table_2}. We can see that both arithmetic means and combined $\chi^2$ measures retain the null hypothesis of uniform randomness across all degrees of freedom, setting the pseudo-RNG apart from our market analogue. We also tested a simplistic pseudo-RNG based on logistic maps with iterative seed draws to confirm the overlapping permutations test's ability to pick up on weak pseudo-RNGs. The sums in particular often barely qualify or fail the test for uniform randomness, highlighting the need for well-tested generators for comparative purposes.

Results for company-separated data closely trace each other for empirical data and pseudo-RNG simulations, with marginally larger means and standard deviations for the former, and with means across both experiments maintaining the null hypothesis of uniformly-random data for all window sizes. The same holds true for combined $\chi^2$ measures once the subset of high-impact contributors are removed, as described in Section~\ref{sec:tests_of_monthly_information_incorporation}. The proportion of statistically significant measures is also approximately the same for company-level Nasdaq data and pseudo-RNG simulations.

Contrary to that, year-separated experiments differ prominently between empirical and simulated data; means and standard deviations taken over annual measures are considerably larger than for the pseudo-RNG output. The latter also features a proportion of statistically significant results similar to the company-separated simulation, whereas the empirical dataset consists mostly of instances satisfying the criterion for inefficiency. As shown in Section~\ref{sec:empirical_experiments_and_results}, a small number of stocks not filtered out in a year-by-year analysis drives much of these large values, although that specific impact does not explain the stark variability between years, with mostly years before the recent global financial crisis qualifying for market efficiency for some of the window sizes.

\section{Discussion}
\label{sec:discussion}

We have shown that there are significant year-to-year changes in exchange-wide efficiency, as well as an overall inefficiency in aggregated annual data. We also find that individual stocks of Nasdaq-listed companies are efficient in aggregate when taking small subsets of inefficient outliers into account, which offers a partial explanation of annual variability, and that stocks follow approximately the same level of uniformly-random assessment as well-tested pseudo-RNGs. The last point is especially relevant to annual anomalies, as they can both be driven by inefficient subsets and be resolved in terms of efficiency over longer time frames, placing the market in a state of overall efficiency at a larger scale. While we have adjusted monthly and daily close prices for splits and dividends as described in Section~\ref{sec:data_preprocessing_and_considerations}, experiments with unadjusted raw prices yield almost identical results. One area that warrants a closer look is the compatibility of our results with the concept of market efficiency in general.

\citet{Samuelson1973} formalises a random walk model of market efficiency, demonstrating both the martingale property of such a model and its allowance for subsets of market participants, too small to affect prices appreciably, to systematically realise excess returns. This is, of course, especially relevant in terms of stronger forms of market efficiency, for both fundamental analysis as permitted under the semi-strong EMH and strong-form insider trading. It also means that market data that is transformed into binarised returns can still contain hidden inefficiencies exploited by small pools of capital, drawing a line between the model of theoretical efficiency and the leeway that practical implementations allow. The result in terms of uniform randomness in empirical data, in both cases and as far as statistical analyses go, is the same.

This bears similarity to two different proposals in the literature; self-destruction of predictability as described by \citet{Timmermann2004}, which posits that anomaly exploitability decays due to a time-limited presence or public dissemination of the anomaly, and the adaptive market hypothesis by \citet{Lo2004}, which attempts to reconcile market efficiency with behavioural economics through adaption processes in changing market environments.

Research on the latter mentioned in Section~\ref{sec:information_efficiency_in_financial_markets} generally focusses on foreign investment, market microstructure factors, and calendar effects. In contrast, we propose an additional technological perspective on the market environment and market actor adaptability. New developments, be it in terms of computing resources or methodology, do not push adopted approaches to market participation out due to the transitory nature of exploitable anomalies or widespread adoption following publication.

Instead, the process is a result of a technological arms race that renders prior solutions unfit for the changed market environment. An already established and prominent example of this process is the competition in terms of information transmission speeds among high-frequency trading firms. In recent years, the adoption of modern machine learning among financial practitioners, as well as the fast development of new methods in the field, has provided further fuel \citep{Gogas2021}. However, as long as adopted technologies, or satisficing heuristics under the terminology of the adaptive market hypothesis, do not outperform to a degree that renders predecessors ineffective, small pools of capital, akin to a small number of species in an abundant environment, can exploit anomalies in a shared manner.

 The above paragraphs show that for the effects of theoretical market efficiency to occur, at least on a meaningful level and for varying notions of market efficiency, the underlying process can contain complications related to inefficiency. This should not come as a surprise, as the EMH, just like models in other disciplines rooted in the scientific method, is a model with explanatory power that does, necessarily, allow for a certain degree of leeway to remain succinct. Anomalies detected in our experiments are, thus, reconcilable with practical market efficiency.

With regard to financial economics and econometrics, this paper provides valuable insights on the time-dependent variability of weak-form market efficiency, as well as on the role of outliers in the assessment of overarching exchanges and broad indices. Natural follow-ups to this type of investigation are the more fine-grained analyses of flash crashes and financial crises, which \citet{Noakes2016} already start for the impact of the Global Financial Crisis of 2007--2008 in South Africa, the potential for industry sector influences and similar effects that shape the presence and importance of inefficient subsets, and the measurement of differences between exchanges, both in terms of subset-driven inefficiency and annual variance, as well as regarding the impact of exchange volatility on overlapping permutations tests.

Similarly, in the field of market microstructure, the question arises whether the latter differences can be linked to exchange peculiarities such as trading rules, systems, and accessibility of advances in financial technology such as high-frequency trading. While our paper, due to limitations in its scope, follows \citet{Noakes2016} in focussing on a particular exchange, a comparison to other developed markets, for example in Europe, is an interesting follow-up avenue \citep{Borges2010}.

In the same vein, and regarding the mentioned differences in trading environment and technology, emerging markets also warrant further study to extend this area of application \citep{Gregoriou2009}. Lastly, the same approach can be transferred to different types of markets outside of stock exchanges. Here, the efficiency of foreign exchange markets is a prime target for follow-up research as a long-standing topic of interest in the financial literature \citep{Burt1977, Chaboud2014}. As mentioned in Section~\ref{sec:comparison_to_measurements_of_daily_data}, different time scales are an interesting extension to these kinds of studies, and varying efficiencies of foreign exchange markets for varying data granularity is an additional direction for future research.

\section{Conclusion}
\label{sec:conclusion}

This paper builds on and extends a topic that has recently developed in the operational research literature, centred on the application of overlapping permutations tests from the field of random number generation to financial exchanges, to test for informational efficiency in markets. To this end, we go beyond existing research by covering a larger and more recent time frame with longer step sizes, and by splitting our experiments into the company level and year-separated analyses for Nasdaq data.

Our results for company-separated data demonstrate that stocks of individual Nasdaq-listed public companies feature average market efficiency as measured in this study, although this efficiency is only confirmed when omitting a small subset of outliers, which skew the overall assessment towards statistically significant inefficiency for the overall exchange. This has implications for prior research on whole markets and overarching indices, and for hypothesis tests of market efficiency more generally. For daily instead of monthly close prices, the number of outliers is slightly larger, as shorter-term inefficiencies in price behaviour can contribute to the results, and this increase is driven by short patterns spanning only a few days.

When performing the same analysis on year-separated data instead, we find that the same effect applies, but also that assessments vary starkly in their pattern recurrence, which is further confirmed through the distribution of summed counts, and reflects cross-correlations and decreased efficiency during financial crisis scenarios.

For both streams, we perform comparisons to a well-tested pseudo-random number generator and find comparable measures for company-separated data once outliers are removed, while annual analyses differ in their year-to-year variation. We also discuss the implications of theoretical versus practical efficiency for market participants,  arguing for the latter kind of efficiency to allow for adaptive leeway as well as unrealised inefficiencies while maintaining the results implied by financial theory.

Our work contributes to the literature on cross-disciplinary methodology transfers in operational research, applications of cryptographic tools in econometric analyses, the evolution of weak-form inefficiency as an anomaly on volatile exchanges in developed markets, and the broader study of exchange efficiency on the individual company level as well as differences between exchanges and links to market microstructure.

\section*{Acknowledgements}

Special thanks go to Antonia Gieschen, whose comments on the potential role of outliers have made the analyses in this paper more complete, as well as Gbenga Ibikunle for previous conversations on the intricacies of testing for market efficiency. We also wish to express our gratitude to the two anonymous reviewers whose comments helped to improve this paper.

%\section*{References}

\bibliographystyle{apalike}
\bibliography{rng_market_efficiency.bib}

\end{document}